\documentstyle[12pt,aasms4]{article}

\begin{document}
  
\title{
The energy spectrum observed by the AGASA experiment and the 
spatial distribution of the sources of ultra-high energy cosmic rays
}

\author{Gustavo A. Medina-Tanco$^{1,2}$}

\affil{ 1.
Instituto Astron\^omico e Geof\'{\i}sico, University of S\~ao Paulo, Brasil \\
gustavo@iagusp.usp.br
}

\affil{ 2. Dept. of Physics and Astronomy, University of Leeds, 
Leeds LS2 9JT, UK}

\singlespace

\begin{abstract} 

Seven and a half years of continuous monitoring of giant air showers triggered 
by ultra high-energy cosmic rays have been recently summarized by the AGASA 
collaboration. The resulting energy spectrum indicates clearly that the cosmic 
ray spectrum extends well beyond the Greisen-Zatsepin-Kuzmin (GZK) cut-off at 
$\sim 5 \times 10^{19}$ eV. Furthermore, despite the small number statistics 
involved, some structure in the spectrum may be emerging. Using numerical 
simulations, it is demonstrated in the present work that these features are 
consistent with a spatial distribution of sources that follows the distribution 
of luminous matter in the local Universe. Therefore, from this point of view, 
there is no need for a second high-energy component of cosmic rays dominating 
the spectrum beyond the GZK cut-off. 


\end{abstract}

\clearpage
                              
\section{Introduction}

Several mechanisms have been proposed for the acceleration of UHECR. At the 
most general level, they can be classified into two large groups: bottom-up 
and top-down mechanisms. Bottom-up mechanisms, although more conservative, 
imply the stretching of rather well known acceleration processes to their 
theoretical limits (and sometimes beyond). They involve particle acceleration 
in the accretion flows of cosmological structures (e.g., Norman, Melrose and 
Atcherberg 1995, Kang, Ryu and Jones 1996), galaxy collisions (Cesarsky and 
Ptuskin 1993, Al-Dargazelli et al. 1997, but see Jones 1998), galactic wind 
shocks (Jokipii and Morfill 1987), pulsars (Hillas 1984, Shemi 1995), active 
galactic nuclei (Biermann and Strittmatter 1987), powerful radio galaxies 
(Rawlings and Saunders 1991, Biermann 1998), gamma ray bursts (Vietri 1995, 
1998, Waxmann 1995, but see Stanev, Schaefer and Watson 1996), etc.. Top-down 
mechanisms, on the other hand, escape from the acceleration problems at the 
expense of exoticism. They already form the particles at high energies and 
involve the most interesting Physics, for example: the decay of topological 
defects into superheavy gauge and Higgs bosons, which then decay into high 
energy neutrinos, gamma rays and nucleons with energies up to the GUT scale 
($\sim 10^{25}$ eV) (e.g., Bhattacharge, Hill and Schramm 1992, Sigl, Schramm 
and Bhattacharge 1994, Berezinsky, Kachelrie and Vilenkin 1997, Berezinsky 
1998, Birkel and Sarkar 1998), high energy neutrino annihilation on relic 
neutrinos (Waxmann 1998), etc..

In general, the distribution of sources of UHECR particles in bottom-up 
mechanisms should be related to the distribution of luminous matter in 
the Universe. In contrast, for top-down mechanisms, an isotropic 
distribution of sources should be expected in most of the models 
(c.f., Hillas 1998, Dubovsky and Tinyakov 1998). Hence the importance 
of distinguishing observationally between these two scenarios.

A possible correlation between compact radio quasars and the five 
most energetic UHECR has already been proposed by Farrar and 
Biermann (1998).
Furthermore, 
the clusters of events observed by AGASA (Hayashida et al 1996) are 
consistent with UHECR production regions at distances of the order of 
$\sim 30$ Mpc, for an intervening IGMF $\sim 10^{-10}$ to $10^{-9}$ Gauss 
(Medina Tanco 1998a). Local maxima in the galaxy density distribution are 
located at those positions. This can be viewed as a point in favor of the 
hypothesis that the UHECR sources are distributed in the same way as the 
luminous matter in the local Universe does. Furthermore, it could naturally 
explain the extension of the UHECR spectrum beyond the GZK cut-off hinted 
by extreme high energy events of Volcano Ranch (Linsley 1963, 1978), Haverah 
Park (Watson 1991, Lawrence, Reid and Watson 1991), Fly's Eye (Bird et al. 
1995) and AGASA (Hayashida et al., 1994), and recently confirmed by the 
latter experiment (Takeda et al. 1998, Nagano 1998).

In the following sections we use numerical simulations to assess both, the 
statistical significance of the AGASA result (Takeda et al. 1998) at the very 
end of the energy spectrum, and the degree to which it is compatible with a 
non-homogeneous distribution of sources that follows closely the spatial 
distribution of luminous matter in the nearby Universe. The possibility of 
solving the puzzle in few years of integration with the next generation of 
large area ($10^{3}$ km$^{2}$) experiments is exemplified through the 
Southern site of the Auger observatory.

\section{Numerical approach and discussion of results}

Energy losses due to photo-pion production in interactions with the 
cosmic microwave background, should lead to the formation of a bump in 
the spectrum beyond $5 \times 10^{19}$ eV, followed by the GZK cut-off 
(Greisen 1966, Zatsepin and Kuzmin 1966) at higher energies. 
The existence and 
exact position of these spectral features depends on the spatial 
distribution of the sources, their cosmological evolution and injection 
spectrum at the sources (Berezinsky and Grigor'eva 1988). Nevertheless, 
both bump and cut-off tend to smooth away for predominantly nearby 
sources or strong cosmological evolution. The most natural way to avoid 
the GZK cut-off is by invoking either top-down mechanisms or the 
existence of relatively very near (compared with the UHECR mean 
free path) sources.

The spectrum calculated by Yoshida and Teshima (1993) for an isotropic, 
homogeneous distribution of cosmic ray sources, and shown superimposed 
on the observed AGASA spectrum in Takeda et al (1998), seems unable to 
explain the extension of the UHECR spectrum beyond $10^{20}$ eV. It is 
not clear, however, whether the available data ($461$ events for 
$E > 10^{19}$ eV, and only $6$ events for $E > 10^{20}$ eV) is 
sufficient to support any conjecture about the actual shape of 
the spectrum above $10^{20}$ eV. Furthermore, it is the nearby sources 
that are expected to be responsible for this region of the spectrum 
and their distribution is far from isotropic or homogeneous. 
Therefore, it is not clear either what is the influence that 
the differential exposure in declination, peculiar to the AGASA 
experiment, has on the deduced spectral shape at the highest energies. 

To analyze the effects of the previously mentioned factors on the 
observed energy spectrum, two different sets of simulations are 
discussed here. 

As a check on the simulations, the energy spectrum by Yoshida and 
Teshima (1993) was reproduced using a homogeneous distribution of 
sources from $z=0$ to $z=0.1$, including adiabatic energy losses 
due to redshift, and pair production and photo-pion production due 
to interactions with the cosmic microwave background radiation (CMBR)
in a Friedmann-Robertson-Walker metric. Furthermore, a fiducial 
intergalactic magnetic field (IGMF), characterized by an intensity 
$B_{IGMF} = 10^{-9}$ G and a correlation length $L_{c} = 1$ Mpc 
(cf., Kronberg, 1994), was also included. The IGMF was assumed uniform 
inside cells of size $L_{c}$ and randomly oriented with respect to 
adjacent cells (Medina Tanco et al 1997). The IGMF component was 
neglected in the original work of Yoshida and Teshima (1993). 
Individual sources were treated as standard candles supplying 
the same luminosity in UHECR protons above $10^{19}$ eV. The 
injected spectrum was a power law, $dN/dE \propto E^{-\nu}$ , 
with $\nu = 3$ above the latter threshold. From the $\sim 10^{7}$ 
particles output by the simulation and arriving isotropically in 
right ascension and declination, one hundred samples were extracted, 
with the same distribution in declination as the quoted exposure of 
AGASA (Uchihori et al. 1996). The determination of the arrival 
energy of protons was performed assuming an error of 20\% 
(energy-independent Gaussian distribution), typical of AGASA 
(e.g., Yoshida and Dai 1998). 

Similarly, the same bin and number of events above $10^{19}$ eV 
($461$ protons) as in the AGASA paper (Takeda et al, 1998) were 
used here for the calculation of the spectra.

The resulting predicted spectrum is shown in figure 1, where the 
different shades indicate 63\% and 95\% confidence levels, i.e., 
the region in the $E^{3} \times dJ/dE$ vs. $E$ space where $63$\% 
and $95$\% of the spectra fell respectively. It can be seen that, 
as predicted by Yoshida and Teshima (1993), the model is able to 
fit the observed AGASA spectrum quite well up to $\sim 10^{20}$ 
eV. The introduction of the IGMF does not make appreciable changes. 
At higher energies, however, AGASA observations seem unaccountable 
by the homogeneous approximation, even when the quoted errors are 
considered.

The distribution of luminous matter in scales comparable with a 
few mean free paths of UHECR protons in the CMBR (i.e., tens of Mpc) 
is, nevertheless, far from homogeneous. Therefore, given the 
relatively small mean free path of protons above $10^{20}$ eV, it 
should be expected that the local distribution galaxies plays a 
key role in determining the shape of the UHECR spectrum if the 
sources of the particles have the same spatial distribution as 
the luminous matter.

The second set of simulations is intended to address the latter 
problem. In figure 2, the number of galaxies inside shells of 
constant thickness in redshift, $\Delta z = 0.001$, are shown as 
a function of $z$ for the latest release (version of Jul 27, 1998) 
of the CfA Redshift Catalogue (Huchra et al 1992). Also shown in 
the same figure is a homogeneous, isotropic distribution of sources. 
The normalization of the latter is such that both distribution 
enclose the same number of galaxies inside $r_{0} = 100$ Mpc. 
The observed distribution of galaxies shows an excess for $r < 60$ 
Mpc compared to the homogeneous distribution. Between $r \sim 60$ and 
$r \sim 100$ Mpc both distributions increase with the same slope. This 
suggests that the approximation of homogeneity begins to be valid 
beyond $r \sim 60$ Mpc and that the actual distribution of galaxies 
is reasonably well sampled (even if obviously incomplete) up to 
$r \sim 100$ Mpc $= r_{0}$. Farther away the slope of the observed 
distribution changes abruptly, very likely due to the predominance 
of bias effects. 

The approximation adopted here is, therefore, that the distribution 
of luminous matter at $r < r_{0} = 100$ Mpc is well described by the 
CfA catalog, while the homogeneous approximation holds outside that 
volume. The previously described simulation scheme is used for the 
distant sources in the homogeneous region, while the actual 
distribution of galaxies is used for the UHECR sources nearer 
than $100$ Mpc.  Additionally, in the latter case, the same 
procedure as in Medina Tanco (1997, 1998a) is used in the description 
of the IGMF: a cell-like spatial structure, with cell size given by 
the correlation length, $L_{c} \propto B_{IGMF}^{-2}(r)$. 
The intensity of the IGMF, in turn, scales with luminous matter 
density, $\rho_{gal}$ as $B_{IGMF} \propto \rho_{gal}^{0.3}(r)$ 
(e.g., Vall\'ee 1997) and the observed IGMF value at the Virgo 
cluster ($\sim 10^{-7}$ G, Arp 1988) is used as the normalization 
condition. 

The resultant spectrum is obtained by combining both contributions, 
from nearby and distant sources respectively, after taking into 
account the complicating fact that our knowledge of the distribution 
of galaxies is not uniform over the celestial sphere (e.g., obscuring 
by dust over the galactic plane). The results, particularized for the 
AGASA experiment (i.e., same declination exposure and energy error, 
as well as number of events and bin size), are shown in figure 3. 
It can be seen that, when the actual distribution of galaxies is 
taken into account, the $63$\% confidence spectrum is able to fit 
all the data if the corresponding experimental errors are considered. 
Consequently, the UHECR spectrum observed by the AGASA experiment is, 
given the available data, compatible with a distribution of cosmic ray 
sources that follows the distribution of luminous matter in the Universe. 
The latter is true up to the highest energies observed so far. Clearly, 
more data is needed before the hypothesis can be falsified.

There is hope, however, for a solution in the relatively near future. 
The same calculations have been performed for the first three years 
of operation of the future Southern site of the Auger experiment. 
The appropriate dependence of exposure on declination was used 
(A. Watson, private communication), and the expected number of 
events, i.e., $9075$ (Auger Design Report, 1997). The results 
are given in figure 4, superimposed with the present AGASA spectrum 
and its previously calculated uncertainty.

Finally, a word of caution should be given regarding the possibility 
of large scale structuring of the IGMF (Ryu, Kang and Biermann 1998). 
The effects that this could have on UHECR propagation have been 
discussed extensively in Medina Tanco (1998b). Unfortunately, it 
is not possible to state undoubtedly in which direction this would 
influence the resulting particle spectrum without knowing the exact 
topology of the IGMF and location of nearby sources with respect to 
the field.

\section{Conclusions}

From the previous analysis it can be concluded that, given the 
low number of events detected by the AGASA experiment so far with 
$E > 10^{19}$ eV, the observed UHECR spectrum is consistent with a 
spatial distribution of sources that follows the luminous matter 
distribution in the nearby Universe. In the latter approach a single 
power-law injection energy-spectrum is assumed, extending up to the 
highest observed energies beyond the GZK cut-off. Therefore, based 
on the observational uncertainties at present, there is no need for 
a second UHECR component responsible for the events observed above 
the nominal GZK cut-off. 

Three years of integration by the future Southern site of the Auger 
observatory should suffice to decide whether the spatial distribution 
of UHECR sources is the same of the nearby luminous matter or not. 

I am very grateful to Alan Watson and Michael Hillas for valuable 
comments and interesting discussions, and to the High-Energy 
Astrophysics group of the University of Leeds for its kind 
hospitality. This work was partially supported by the Brazilian 
agency FAPESP.



\newpage

\noindent
{\bf Figure Captions}

\bigskip

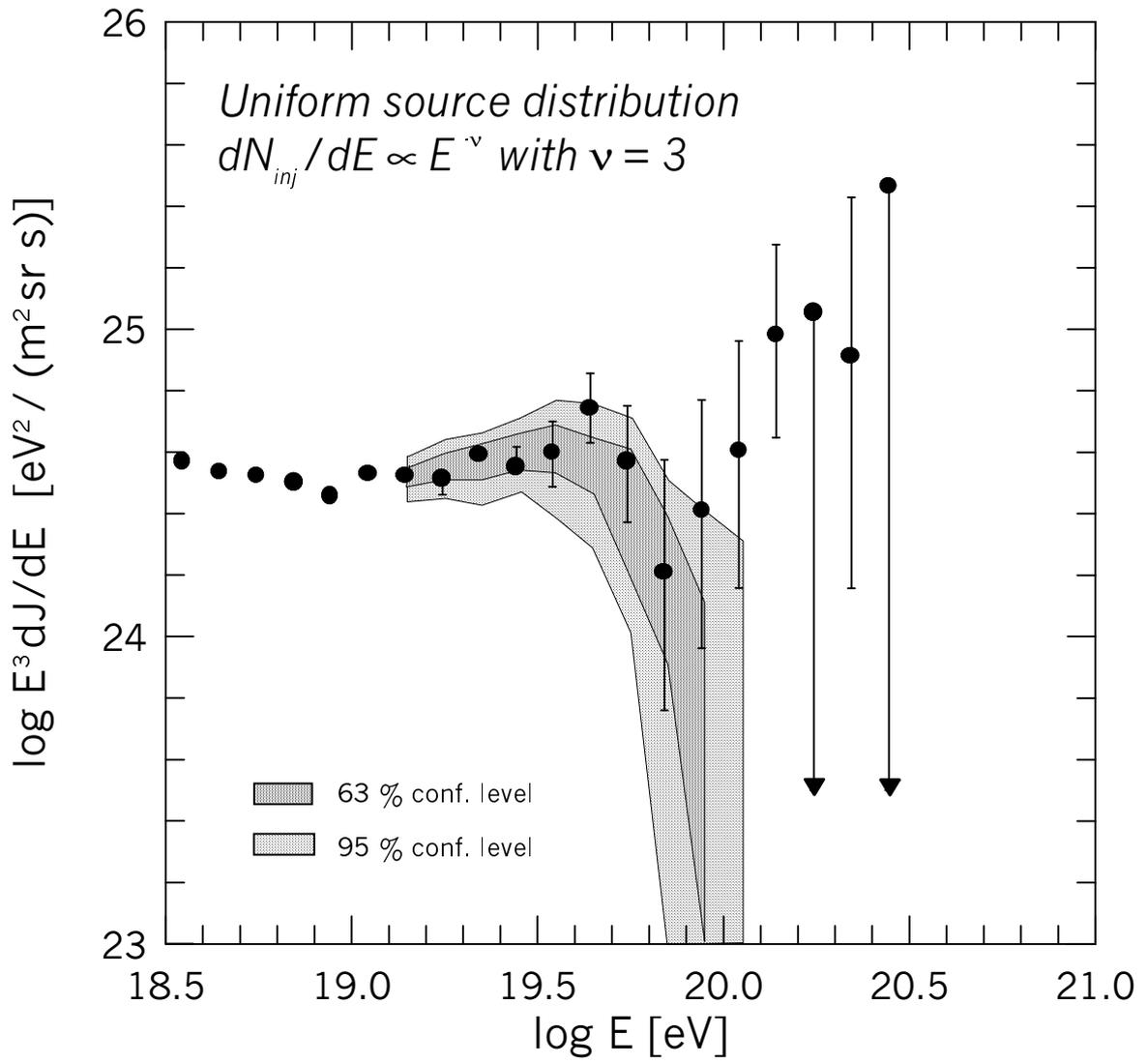
{\bf Figure 1:} : Spectrum from a uniform distribution of 
sources superimposed on the observed AGASA spectrum (adapted 
from Takeda et al. 1998). The same total number of events, 
energy determination error and exposure as a function of 
declination as AGASA's are used. Error bars in the AGASA
spectrum represent the Poisson upper and lower limits at 68\%
and arrows are 98\% confidence upper limits.


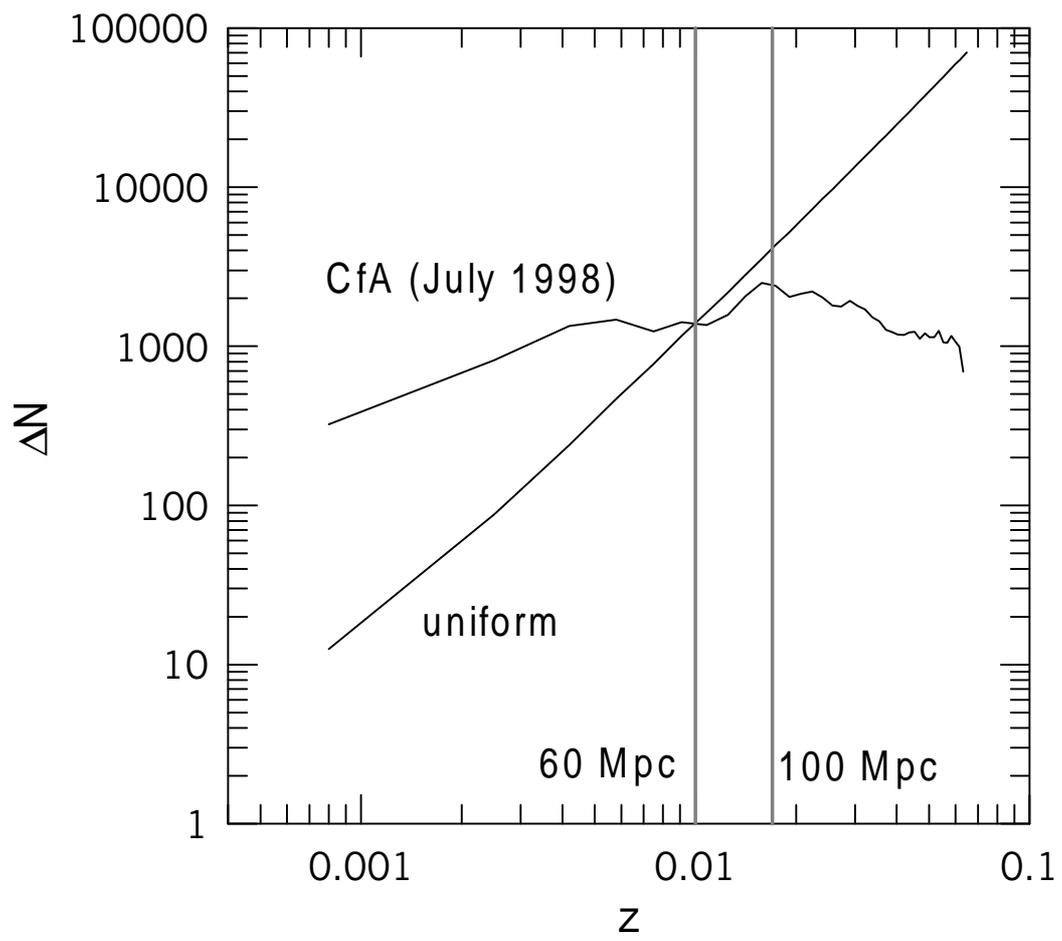
{\bf Figure 2:} Number of galaxies inside shells of constant 
thickness $\Delta z = 0.001$ as a function of redshift z for 
the CfA redshift catalog (July 1998) and a uniform distribution 
of sources. Both distribution were normalized to give the same 
number of sources inside $r_{0} = 100$ Mpc. Note the observed 
excess of galaxies at short distances relative to the uniform 
distribution. Moreover, both curves have the same slope between 
$60$ and $100$ Mpc. This suggests that beyond $60$ Mpc the 
observed distribution behaves approximately as a uniform distribution 
and that the catalog maps correctly the actual distribution of 
galaxies as far as $100$ Mpc. At larger distances the undersampling 
is apparent. (Quoted distances are for $h=0.5$.)


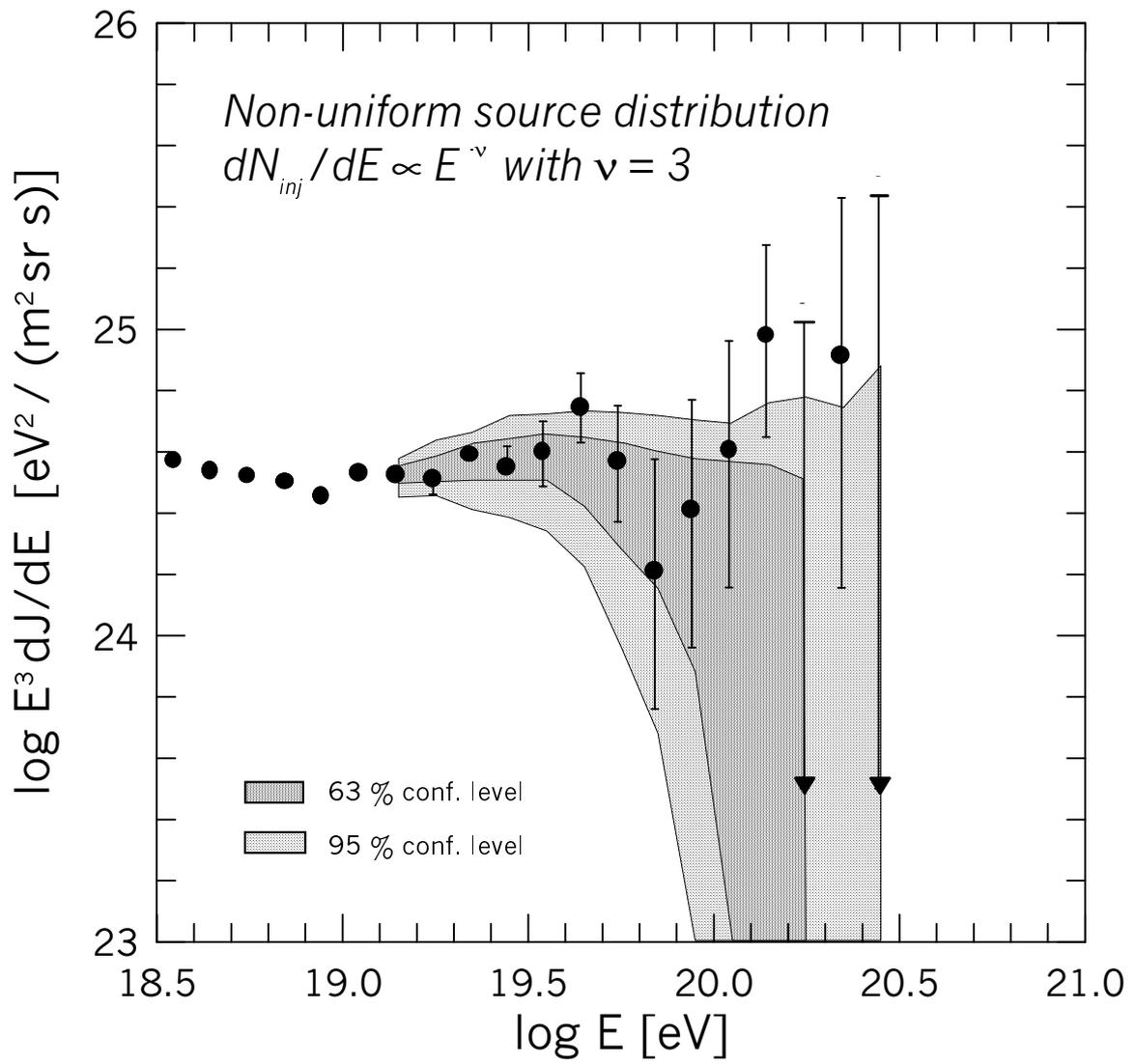
{\bf Figure 3:} Major result of this work. Calculated observable 
spectrum for the same experimental conditions of AGASA, under the 
assumption that the UHECR sources are distributed spatially in the 
same way as the luminous matter in the nearby universe.


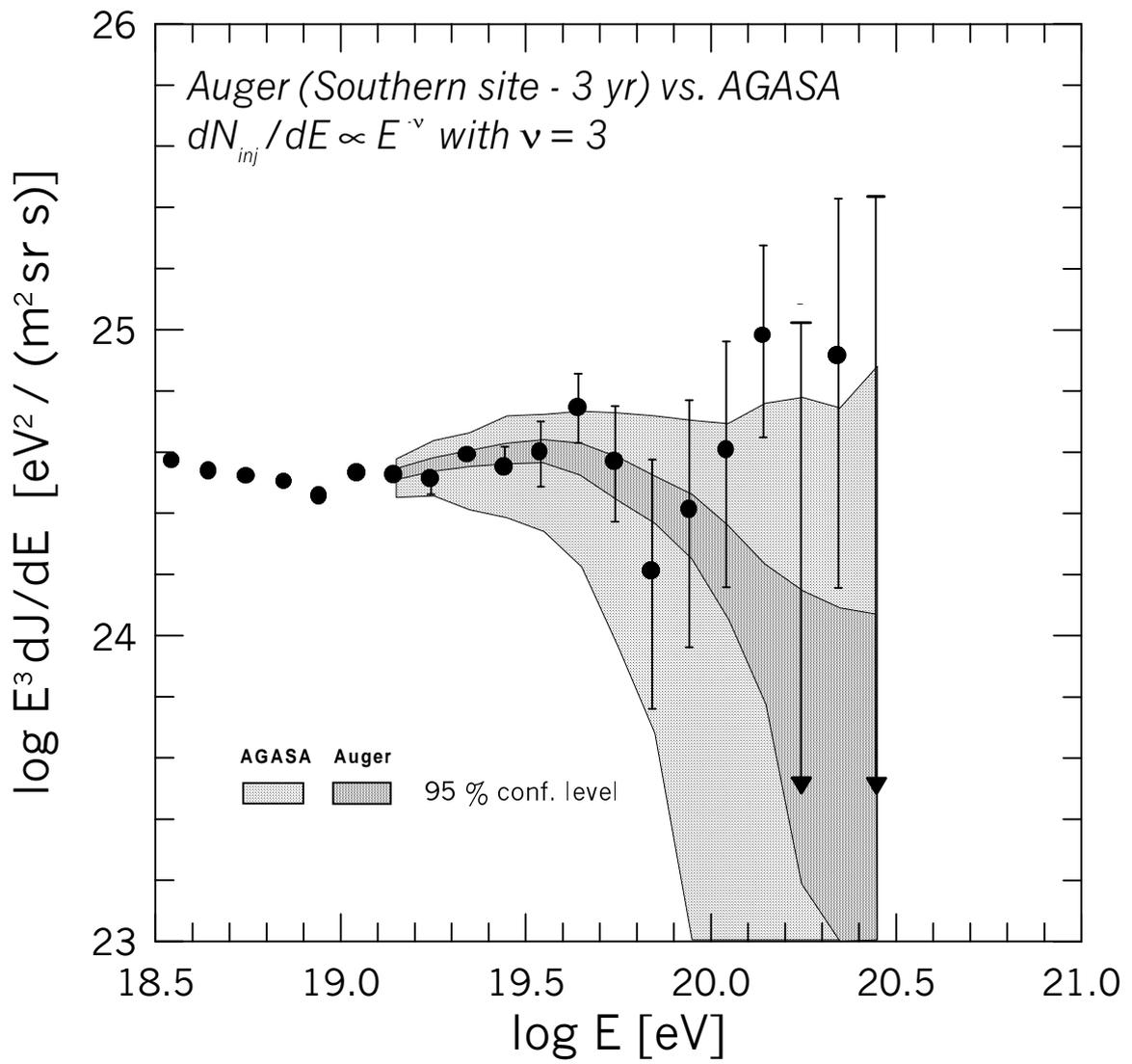
{\bf Figure 4:} The future. The same calculations as in figure 4 
are reproduced for the Southern site of the future Auger experiment. 
The result corresponds to the first three years of observation and is 
compared with the results for the present data of the AGASA experiment. 
Only $95$\% confidence spectra are shown.

\end{document}